# Role of ion migrations and structural reorganisations in femtosecond laser direct-written chalcogenide glass waveguides

Thomas Gretzinger[1,2], Toney Teddy Fernandez[2,*], Simon Gross[1,2], Alexander Arriola[1,2] and Michael Withford[1,2]

[1]*Centre for Ultrahigh bandwidth Devices for Optical Systems (CUDOS),*
[2]*MQ Photonics Research Centre, Department of Physics and Astronomy, Macquarie University, NSW, 2109, Australia*
*toney.teddyfernandez@mq.edu.au

Formation of femtosecond laser direct-written positive refractive index waveguides in Gallium Lanthanum Sulphide (GLS) glass is explained for the first time. Evidence of structural changes and ion migration are presented using Raman spectroscopy and electron probe microanalysis (EPMA), respectively. 2-D Raman spectra maps reveal a peak shift and full-width at half maximum variations in the symmetric vibrations of $GaS_4$ main band. For the first time, the 2D map of the boson band was successfully used to identify and understand the material densification profile in a high refractive index glass waveguide. Finally EPMA provided the evidence of ion migration due to sulphur and the observation of an anion ($S^{2-}$) migration causing material modification is also reported for the first time.

Chalcogenide glasses are a promising platform for integrated optic devices operating in the mid-infrared wavelength region(1). Gallium Lanthanum Sulphide, otherwise known as GLS glass, is one of the most important material available for photonic applications. Several promising applications like astrophotonics, high density memories, supercontinuum generation will be fulfilled by chalcogenide glass built devices(2). Optimising such devices based on femtosecond laser direct-written waveguides requires detailed understanding of the waveguide formation. This paper pin points the exact structural modification in the material upon irradiation with femtosecond laser pulses. Raman spectroscopy and EPMA using wavelength dispersive X-ray spectroscopy (WDS) were the major tools used in this work. Previous reports in other glasses like phosphates, tellurites and borosilicates have helped to understand the waveguide formation from its basic elemental units due to migration of constituent ions(3). This is the first report to provide detailed evidence of the structural modification and ion migration in chalcogenide glasses(4, 5).

The GLS glass samples were kindly provided by ChG Southampton Ltd. in sizes of 10×50×1 mm with all top and bottom surface polished. For inscribing waveguides in the thermal regime, i.e. cumulative heating, a 5.1 MHz Ti:sapphire chirped pulse femtosecond oscillator (Femtosource XL500, Femtolasers GmbH) was used. The laser emits up to 550 nJ pulses at 800 nm wavelength with a pulse duration of <50 fs. The circularly polarised pulses were directed by an optical setup to an Olympus Plan N 100× oil immersion microscope objective (NA ≈ 1.25, $NA_{eff}$ ≈ 0.9). Oil immersion reduces the refractive index mismatch and thus mitigates spherical aberrations compared to air objectives(6). The samples were placed on a set of 3-axis computer controlled Aerotech air-bearing stages. To achieve low loss waveguides (propagation loss ~ 0.22 dB/cm ±0.02) at 4 µm wavelength, three modifications were inscribed next to each other at a writing depth of 180 µm. The sample was translated at a feedrate of 100 mm/min with a pulse energy of 13 nJ. A cross sectional DIC image of the inscribed single-track waveguide is shown in **Figure 1a** and the final waveguide structure consisting of three single tracks stacked laterally in **Figure 1b.** The guided single-mode, in the triplet structure, at 4 µm wavelength is shown in **Figure 1c**. The 'core' and 'shell' regions marked in Figure 1a will be the terminologies used throughout in this manuscript for easy identification while discussing them.

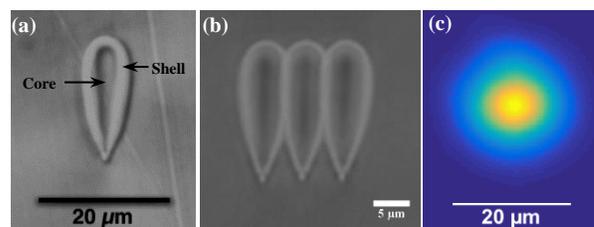

**Figure 1** DIC image of **(a)** single and **(b)** triple waveguide structure. **(b)** Guided mode at 4 µm wavelength.

The Raman spectroscopy was carried out using a Horiba LabRam HR Evolution confocal Micro Laser Raman spectrometer. The spectrometer was operated in confocal mode with a 633 nm laser for excitation. A 100× (0.9 NA) objective was used to map the waveguides. The objective provided a spot size <1 µm and the spectra were acquired at a resolution of 1 cm$^{-1}$. The EPMA was carried out using both energy dispersive X-ray spectroscopy (EDX) and WDS attached to a Zeiss EVO MA15 SEM and Cameca SX100 SEM, respectively.

Both low and high repetition rate femtosecond laser formed waveguides are reported in GLS glass(4, 5, 7-9), of which the heat accumulated waveguides are of greater interest due to the rapid fabrication times and higher refractive index contrast(8). Both these factors are key when designing waveguides for the mid-infrared region which requires relatively larger waveguide structures and/or higher refractive index contrast for the best mode-matching conditions. More details about the waveguide fabrication presented in this paper can be found elsewhere(10).

Neither the 2D refractive index profile nor the structural integrity of the chalcogenide waveguides in GLS glass are clearly understood to date. Measuring the refractive index profile of a waveguide inscribed in a high refractive index glass (>1.7) is still quite challenging due to lack of feasible and reliable characterisation techniques. Previous attempts to measure the refractive index profile of a similar GLS glass waveguide structures using quantitative phase microscopy (QPM) provided a 1-D quantitative result(5). Further attempts using ellipsometric measurements were also unsuccessful(11). The lack of quantitative and qualitative information about the 2D refractive index distribution makes deterministic waveguide design using computer software challenging and thus slows progress. In this work we have decided to use the information available from the excess density of states (DOS) boson peak (BP) in the Raman spectrum which is sensitive to local density fluctuations(12). The DOS of the amorphous system (BP) is also considered to be a modification of the crystalline DOS (Van Hove singularity) due to a random fluctuation of force constants. The boson peak shows up in any amorphous material regardless of its constituents or stoichiometry and is observed in the low frequency region between $0 - 100$ cm$^{-1}$. Their origin is a well debated topic and is mostly controversial, here we make use of its response to the physical properties of the material. Hence a 2-D Raman mapping of the boson band around the waveguide region is ideal to deduce qualitative information of the refractive index profile. Adding to this picture, the presence of a BP in an amorphous material is an indication of a more elastic matrix as opposed to a weak fragile structure. The vibrational wavenumber dependence on the boson peak and intensity was experimentally demonstrated by Inamura et. al (13). They demonstrated that as the densification happens the low energy modes of the boson peak get suppressed shifting the peak to higher wavenumbers.

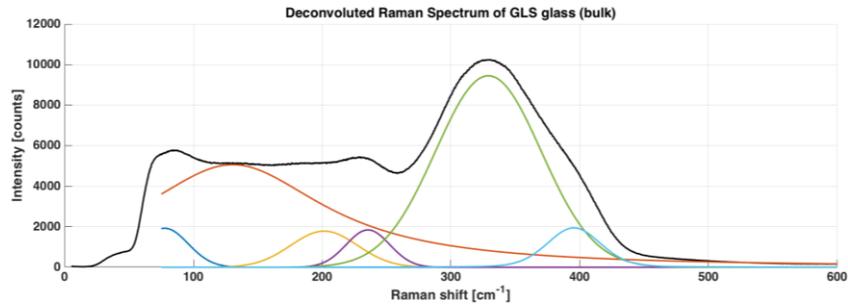

**Figure 2** De-convolved Raman spectra of bulk GLS glass

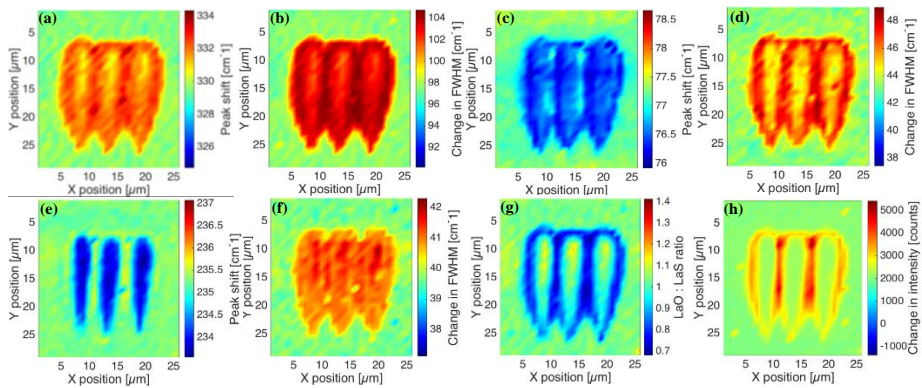

**Figure 3a-h** 2D Raman mapping of the waveguide

The Raman spectrum collected from the bulk glass is shown in the **Figure 2.** The peaks were identified and de-convolved by pseudo-Voigt curve fitting. The identified peaks were the following: 77.3 cm$^{-1}$ (boson peak)(14), 131.0 cm$^{-1}$ (GaS$_4$ asymmetric stretching mode), 201.5 cm$^{-1}$(La-S bending mode), 235.2 cm$^{-1}$(Ga$_2$S$_3$ distorted crystalline like units) 329.5 cm$^{-1}$ (GaS$_4$ symmetric stretching mode), 395.7 cm$^{-1}$(La-O bending mode)(15, 16). As the DIC microscope image in Figure 1a shows, each laser direct-written modification has a central inverted tear drop shaped core like structure (**core**) with a concentric bright zone around it (**shell**). The GaS4 symmetric stretching vibration was the strongest band and the 2-D mapping (**Figure 3a**) demonstrated a 0.5 cm$^{-1}$ and 2.5 cm$^{-1}$ shift towards the higher wavenumber in the core and shell respectively. Both these regions simultaneously increased their vibration bandwidth (**Figure 3b**) by 5.0 and 4.0 cm$^{-1}$, respectively. This indicates a larger distribution of bond lengths and/or bond angles, evidencing a strong structural modification. To bring more clarification to this observation and also to map the local density fluctuations, we made use of the valuable information provided by the boson peak at 78.9 $cm^{-1}$. The 2-D mapping of the boson peak (**Figure 3c**) revealed a larger shift to lower wavenumbers for shell region with a magnitude 1.3 $cm^{-1}$, compared to core. Core region instead features isolated patches of almost no shift or even slightly higher wavenumber than the bulk. The average shift of the boson peak in core is less than 0.5 $cm^{-1}$. A shift to lower wavenumbers indicates low density zones. There is an increase in FWHM of the boson peak (**Figure 3d**) at shell region (4.5 $cm^{-1}$) indicating the activation of lower energy modes in agreement with the reports of Inamura et.al. Similar results were obtained in the report of Boolchand et.al(17) where low frequency floppy modes increase in the vicinity of the boson peak when the concentration of chalcogen components was increased in a chalcogenide glass. This was the first clue of sulphur ion migration as low density with higher refractive index can be attributed to sulphur migration as it possesses higher electronic polarisability than any other constituents in the GLS glass. Presence of structures similar to $Ga_2S_3$ crystals were identified by the peak at 235.3 $cm^{-1}$. The mapping of this peak (**Figure 3e**) revealed a decrease in wavenumber in core and a slight increase in shell region respectively, indicating a species migrating behaviour. The net effect of this migration leads to a broadening (**Figure 3f**) of the $Ga_2S_3$ peak irrespective of core-shell definitive zones. The 2D map with the intensity ratio between the La-S and La-O vibration (**Figure 3g**), normalised with respect to the bulk, revealed an increased signal due to sulphur in shell and increased signal due to oxygen in core. Hence the two regions could be tentatively identified as an oxygen rich zone (GLSO) for core and sulphur rich zone (GLS) for shell. The incoming S atoms due to this migration can break the Ga-S dative covalent bond producing a di-coordinated S bridge with a $La^{3+}$ reception site, which eventually increases the presence of sulphur sites with $S^{2-}$ to $La^{3+}$ coordination number 7.5 and the GLSO region where coordination number will be reduced to 7 for $La^{3+}$ ions due to $O^{2-}$ enrichment (16, 18). In the GLS glass $GaS_4$ units are the glass formers whereas $La_2S_3$ is the network modifier and $La_2O_3$ is the network stabiliser. $La_2O_3$ is added to stabilise the sulphide and oxide negative cavities formed associated to the $GaS_4$ tetrahedron which eventually help the $La^{3+}$ ions to be dissolved into the glass matrix reducing crystallisation and aid glass formation. In the present glass from ChG Southampton Ltd, no Lanthanum oxides were added, but we assume that the oxide impurities could get included from traces of each precursor added for glass formation or from the atmosphere or oxygen being present in the glass via the ever-present hydroxyl groups. GLS glasses deviate from the general tendency (Wemple equation(18)) of increasing the refractive index with decreasing average electronic band gap, and/or increasing electronic oscillator strength a trend that is found in majority of glasses. For glasses previously reported with $70Ga_2S_3\cdot30La_2x_3$ where $x$ is replaced by S (GLS) or O (GLSO), a higher density is reported for GLSO (4.26 g/cm$^3$) than GLS (4.03 g/cm$^3$) and additionally the average electronic band gap is found larger for GLSO. But due to the larger electronic polarisability of $S^{2-}$ ions(19) than $Ga^{3+}$, $La^{3+}$ and $O^{2-}$ ions, the refractive index of GLS is much higher compared to GLSO. This is in agreement with the boson peak shift (Figure 3c) done in this paper to reveal the density fluctuations at the waveguide site. Previous studies based on vitreous silica and other glasses shows that boson peak position shifts to higher frequencies with increasing density(12, 13). In this report, we observe a shift to lower frequency indicating reduced density. We also observe a trend that is previously reported in silica where there is an intensity enhancement at the low density zones (**Figure 3h**) which is generally attributed to the buckling motion in the network associated with the distribution of free volume. This signifies an increase in quasi-periodicity consistent with increased void space due to low density. This observation of higher disorder is congruent to the general increase in FWHM of the main Raman vibration bands viz. $GaS_4$ asymmetric stretching (Figure 3b) and $Ga_2S_3$ distorted crystalline like units (figure 3f). F**igure 3g** indicates an increase in sulphur groups at this low frequency shifted region. Sulphur exhibits a larger polarisability, resulting in a higher refractive index which leads to the formation of guiding region in these laser direct-written structures. This explanation is supported by evidence of compositional changes through EDX mapping. **Figure 4a** shows the backscattered electron (BSE) image which is formed due to the z-contrast of the heavier atomic species. The imaging was done at 15KV, 1 nA excitation on a Zeiss EVO MA15, whereas **Figure 4b** shows a much better resolved BSE image which was acquired using a 15KV, 20 nA excitation on a Cameca SX100. Low overall contrast between the

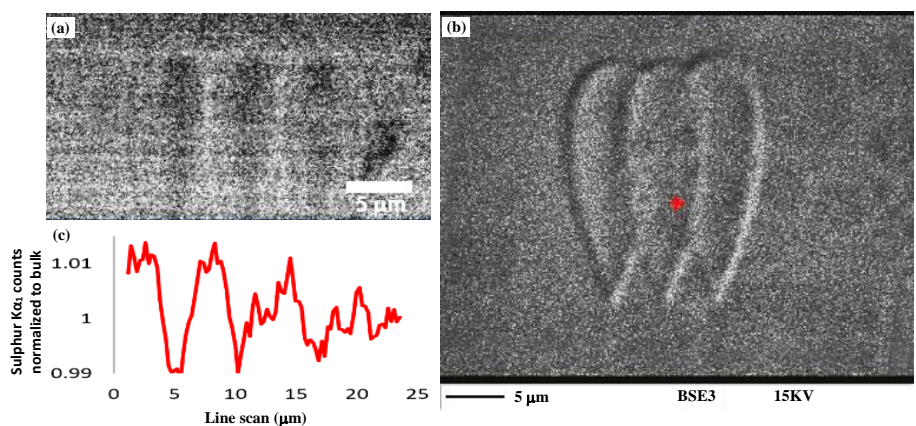

**Figure 4 (a & b)** Backscattered electron image of the waveguide structure. **(c)** The sulphur variation trend with respect to the line scan core and shell zones indicates migration of lighter constituent elements. This result is also in agreement with our previous report(20) on light (regime-1, atomic weight < 31 u) and heavy element (regime-2) ion migration depending on the energy density that is used for waveguide inscription. In this report, the waveguides were inscribed by depositing very low energies and there is no evidence that La and Ga is relocated. This indicates that the waveguide formation is due to regime-1(20). Sulphur is a relatively heavy element compared to oxygen thus resulting in brighter zones in shell region. **Figure 4c** shows the amount of sulphur variation across the waveguide with respect to the bulk. The shell is enriched by sulphur and we expect the migration of oxygen towards core in agreement with Raman spectroscopy. The measurement of oxygen content using electron microscopy is challenging to present and reliably interpret a solid evidence, hence we avoid the same.

In conclusion, the paper explains the waveguide formation in GLS glass with structural modifications due to the femtosecond laser interaction with the bulk and ion migration of sulphur to form regions of high refractive index. Boson peak mapping was applied for the first time to a laser direct-written waveguide to reveal density fluctuations. High refractive index of GLS glass prevents reliable refractive index profiling using readily available techniques, hence we propose this new technique could help profiling structures formed in other amorphous high index materials. The observation of a cation like $S^{2-}$ migrating in a glass is observed for the first time in any material, which indicates the unexplored understanding and achievable potential of femtosecond laser microfabrication down to an atomic scale.

## Acknowledgements

Thomas Gretzinger gratefully acknowledges Dan Hewak and ChG Southampton Ltd for kindly providing the GLS glass samples. An additional credit goes out to Timothy Murphy at the Department of Earth and Planetary Science for the support in operating the Raman, EDX and WDS system.
The work was performed in part at the OptoFab node of the Australian National Fabrication Facility. A company established under the National Collaborative Research Infrastructure Strategy to provide nano- and microfabrication facilities for Australia's researchers.